\documentclass[onecolumn]{aastex}

\usepackage{graphicx}
\usepackage{dcolumn}
\usepackage{bm}
\usepackage{color}
\usepackage{amsfonts}
\usepackage{amssymb}
\usepackage{amsmath}

\input epsf

\begin{document}

\title{\large{\bf Observational Constraints of Modified Chaplygin Gas\\ in RS II Brane}}

\author{{\bf Chayan Ranjit}\altaffilmark{1}}
\author{{\bf Prabir Rudra}\altaffilmark{2}}
\author{{\bf Sujata Kundu}\altaffilmark{3}}

\altaffiltext{1}{{Department of Mathematics, Seacom Engineering
College, Howrah - 711 302, India. Email: chayanranjit@gmail.com}}

\altaffiltext{2}{{Department of Mathematics, Pailan College of
Management and Technology, Bengal Pailan Park, Kolkata-700 104,
India. Email: prudra.math@gmail.com}}

\altaffiltext{3}{{Department of Information Technology, Narula
Institute of Technology, Kolkata-700109,India. Email:
sujatakundu10@gmail.com}}

\date{\today}

\begin{abstract}

FRW universe in RS II braneworld model filled with a combination
of dark matter and dark energy in the form of modified Chaplygin
gas (MCG) is considered. It is known that the equation of state
(EoS) for MCG is a three-variable equation determined by $A$,
$\alpha$ and $B$. The permitted values of these parameters are
determined by the recent astrophysical and cosmological
observational data. Here we present the Hubble parameter in terms
of the observable parameters $\Omega_{m0}$, $\Omega_{x0}$,
$H_{0}$, redshift $z$ and other parameters like $A$, $B$, $C$ and
$\alpha$. From Stern data set (12 points), we have obtained the
bounds of the arbitrary parameters by minimizing the $\chi^{2}$
test. The best-fit values of the parameters are obtained by 66\%,
90\% and 99\% confidence levels. Next due to joint analysis with
BAO and CMB observations, we have also obtained the bounds of the
parameters ($B,C$) by fixing some other parameters $\alpha$ and
$A$. The best fit value of distance modulus $\mu(z)$ is obtained
for the MCG model in RS II brane, and it is concluded that our
model is perfectly consistent with the union2 sample data.
\end{abstract}


\maketitle

\section{\normalsize\bf{Introduction}}

\noindent

Recent cosmological observations of the SNeIa
(\cite{Perlmutter,Perlmutter1,Riess,Riess1}) large scale redshift
surveys (\cite{Bachall,Tedmark}), the measurements of the cosmic
microwave background (CMB) (\cite{Miller,Bennet}) and WMAP
(\cite{Briddle,Spergel}) indicate that our universe is presently
undergoing an accelerated expansion. The observational facts are
not clearly described by the standard big bang cosmology with
perfect fluid. The first suitable candidate which could drive the
acceleration in Einstein's gravity, was the cosmological constant
$\Lambda$ (which has the equation of state $w_{\Lambda}=-1$), but
till now there is no proof of the origin of $\Lambda$. In the
framework of general relativity, different interesting mechanisms
such as loop quantum cosmology \cite{Asthekar}, modified gravity
\cite{Cognola}, higher dimensional phenomena
\cite{Chakraborty2010,Ranjit}, Brans-Dicke theory \cite{Brans},
brane-world model \cite{Gergely} and so on, suggested that some
unknown matters are responsible for accelerating scenario of the
universe and which violates the strong energy conditions, i.e.
$\rho+3p<0$ and which has positive energy density and sufficient
negative pressure, known as dark energy \cite{Paddy,Sahni}. Dark
energy associated with a scalar field is called quintessence
\cite{Peebles}. It is one of the most favored candidate for
producing sufficient negative pressure to drive the cosmic
acceleration, in which the scalar potential of the field dominates
over the kinetic term. In the present cosmic concordance
$\Lambda$CDM model the Universe is formed of $\sim$ 26\% matter
(baryonic + dark matter) and $\sim$ 74\% of a smooth vacuum energy
component. However there is about 0.01\% of thermal CMB component,
but in spite of this, its angular power spectrum of temperature
anisotropies encode important information about the structure
formation process and other cosmic observables.

 If we assume a flat universe and further assume that the only
energy densities present are those corresponding to the
non-relativistic dust-like matter and dark energy, then we need to
know $\Omega_{m}$ of the dust-like matter and $H(z)$ to a very
high accuracy in order to get a handle on $\Omega_{X}$ or $w_{X}$
of the dark energy \cite{Paddy1,Paddy2}. This can be a fairly
strong degeneracy for determining $w_{X}(z)$ from observations.
TONRY data set with the 230 data points \cite{Tonry} alongwith the
23 points from Barris et al \cite{Barris} are valid for $z>0.01$.
Another data set consists of all the 156 points in the ``gold''
sample of Riess et al \cite{Riess1}, which includes the latest
points observed by HST and this covers the redshift range $1 < z <
1.6$. In Einstein's gravity and in the flat model of the FRW
universe, one finds $\Omega_{\Lambda}+\Omega_{m}=1$, which are
currently favoured strongly by CMBR data (for recent WMAP results,
see \cite{Spergel}). In a simple analysis for the most recent
RIESS data set gives a best-fit value of $\Omega_{m}$ to be
$0.31\pm 0.04$. This matches with the value
$\Omega_{m}=0.29^{+0.05}_{-0.03}$ obtained by Riess et al
\cite{Riess}. In comparison, the best-fit $\Omega_{m}$ for flat
models was found to be $0.31\pm 0.08$ \cite{Paddy1}. The flat
concordance $\Lambda$CDM model remains an excellent fit to the
Union2 data with the best-fit constant equation of state parameter
$w=-0.997^{+0.050}_{-0.054}$(stat)$^{+0.077}_{-0.082}$(stat+sys
together) for a flat universe, or
$w=-1.038^{+0.056}_{-0.059}$(stat)$^{+0.093}_{-0.097}$(stat+sys
together) with curvature \cite{Amanullah}. Chaplygin gas is the
more effective candidate of dark energy with equation of state
$p=-B/\rho$ \cite{Kamenshchik} with $B>0$. It has been generalized
to the form $p=-B/\rho^{\alpha}$ \cite{Gorini} and thereafter
modified to the form $p=A\rho-B/\rho^{\alpha}$ \cite{Debnath}. The
MCG best fits with the 3 year WMAP and the SDSS data with the
choice of parameters $A =0.085$ and $\alpha = 1.724$ \cite{Lu}
which are improved constraints than the previous ones $-0.35 < A <
0.025$ \cite{Jun}.

An effective explanation to the late cosmic acceleration can also
be obtained by the modification of Einstein gravity. As a result
various modified gravity theories came into existence.
Brane-gravity is one such modified gravity theory that was
established with the aim of modelling our present day universe in
a better way, and consequently brane cosmology was developed. A
review on brane-gravity and its various applications with special
attention to cosmology is available in \cite{Rubakov, Maartens1,
Brax}. Randall and Sundrum \cite{Randall1, Randall2} proposed a
bulk-brane model to explain the higher dimensional theory,
popularly known as RS II brane model. According to this model we
live in a four dimensional world (called 3-brane, a domain wall)
which is embedded in a 5D space time (bulk). All matter fields are
confined in the brane whereas gravity can only propagate in the
bulk. The consistency of this brane model with the expanding
universe has given immense popularity to this model of late.

Motivated by the previous works of some authors
\cite{Ranjit5,Ranjit8} here we assume the FRW universe in RS II
model filled with the dark matter and the MCG type dark energy.
Our basic idea is to determine the limits of the parameters
involved in the EoS of MCG using the observational data. We
present the Hubble parameter in terms of the observable parameters
$\Omega_{m}$, $\Omega_{x}$ and $H_{0}$ with the redshift $z$. From
Stern data set ($12$ points), the bounds of the arbitrary
parameters is obtained by minimizing the $\chi^{2}$ test. The
best-fit values of the parameters are obtained in $66\%$, $90\%$
and $99\%$ confidence levels. Via a joint analysis with BAO and
CMB observations, we also obtain the bounds and the best fit
values of the parameters ($B,C$) by fixing the other parameters
$A$ and $\alpha$. From the best fit values of distance modulus
$\mu(z)$ for our MCG model in RS II brane, we conclude that our
model is in agreement with the union2 sample data.

The paper is organized as follows: In section 2, the basic
equations and solutions for MCG in RS II braneworld is presented.
The entire data analysis mechanism is given in section 3. Finally
some observational conclusions are drawn in section 4.

\section{Basic Equations and Solutions for MCG in RS II braneworld}

In RS II model the effective equations of motion on the 3-brane
embedded in 5D bulk having $Z_{2}$-symmetry are given by
\cite{Maartens1, Maartens2, Randall2, Shiromizu, Maeda, Sasaki}
\begin{equation}
^{(4)}G_{\mu\nu}=-\Lambda_{4}q_{\mu\nu}+\kappa^{2}_{4}\tau_{\mu\nu}+\kappa^{4}_{5}\Pi_{\mu\nu}-E_{\mu\nu}
\end{equation}
where
\begin{equation}
\kappa^{2}_{4}=\frac{1}{6}~\lambda\kappa^{4}_{5}~,
\end{equation}
\begin{equation}
\Lambda_{4}=\frac{1}{2}~\kappa^{2}_{5}\left(\Lambda_{5}+\frac{1}{6}~\kappa^{2}_{5}\lambda^{2}\right)
\end{equation}
and
\begin{equation}
\Pi_{\mu\nu}=-\frac{1}{4}~\tau_{\mu\alpha}\tau^{\alpha}_{\nu}+\frac{1}{12}~\tau\tau_{\mu\nu}+\frac{1}{8}~
q_{\mu\nu}\tau_{\alpha\beta}\tau^{\alpha\beta}-\frac{1}{24}~q_{\mu\nu}\tau^{2}
\end{equation}
and $E_{\mu\nu}$ is the electric part of the 5D Weyl tensor. Here
$\kappa_{5},~\Lambda_{5},~\tau_{\mu\nu}$ and $\Lambda_{4}$ are
respectively the 5D gravitational coupling constant, 5D
cosmological constant, the brane tension (vacuum energy), brane
energy-momentum tensor and effective 4D cosmological constant. The
explicit form of the above modified Einstein equations in flat
universe are
\begin{equation}
3H^{2}=\Lambda_{4}+\kappa^{2}_{4}\rho+\frac{\kappa^{2}_{4}}{2\lambda}~\rho^{2}+\frac{6}{\lambda
\kappa^{2}_{4}}\cal{U}
\end{equation}
and
\begin{equation}
2\dot{H}+3H^{2}=\Lambda_{4}-\kappa^{2}_{4}p-\frac{\kappa^{2}_{4}}{2\lambda}~\rho
p-\frac{\kappa^{2}_{4}}{2\lambda}~\rho^{2}-\frac{2}{\lambda
\kappa^{2}_{4}}\cal{U}
\end{equation}
The dark radiation $\cal{U}$ obeys
\begin{equation}
\dot{\cal U}+4H{\cal U}=0
\end{equation}
Here $\rho=\rho_{x}+\rho_{m}$ and $p=p_{x}+p_{m}$, where
$\rho_{m}$ and $p_{m}$ are the energy density and pressure of the
dark matter with the equation of state given by $p_{m} =
\omega_{m}\rho_{m}$ and $\rho_{x}$, $p_{x}$ are respectively the
energy density and pressure contribution of some dark energy. Here
we consider an universe filled with Modified Chaplygin Gas (MCG).
The equation of state(EOS) of MCG is given by
\begin{equation}
p_{x} = A\rho_{x}- \frac{B}{\rho_{x}^{\alpha}},~~ B >0 ,~~0\leq
\alpha \leq 1
\end{equation}
We also consider the dark matter and and the dark energy are
separately conserved and the conservation equations of dark matter
and dark energy (MCG) are given by
\begin{equation}
\dot{\rho}_{m}+3H(\rho_{m}+p_{m})=0
\end{equation}
and
\begin{equation}
\dot{\rho}_{x}+3H(\rho_{x}+p_{x})=0
\end{equation}
From first conservation equation (9) we have the solution of
$\rho_{m}$ as
\begin{equation}
\rho_{m}=\rho_{m0}(1+z)^{3(1+\omega_{m})}
\end{equation}
From the conservation equation (10) we have the solution of the
energy density as
\begin{equation}
\rho_{x}=\left[\frac{B}{A+1}+C(1+z)^{3(\alpha+1)(A+1)}\right]^{\frac{1}{\alpha+1}}
\end{equation}
where $C$ is the integrating constant, $z=\frac{1}{a}-1$ is the
cosmological redshift (choosing $a_{0}=1$) and the first constant
term can be interpreted as the contribution of dark energy. So the
above equation can be written as
\begin{equation}
\rho_{x}=\rho_{x0}\left[\frac{B}{(1+A)C+B}+\frac{(1+A)C}{(1+A)C+B}(1+z)^{3(\alpha+1)(A+1)}\right]^{\frac{1}{\alpha+1}}
\end{equation}
where $\rho_{x0}$ is the present value of the dark energy density.

In the next section, we shall investigate some bounds of the
parameters in RS II brane with the assumptions that $\Lambda_{4}$
=$\cal U$ = $0$ (i.e., in absence of cosmological constant and
dark radiation) by observational data fitting. The parameters are
determined by $H(z)$-$z$ (Stern), BAO and CMB data analysis
\cite{Wu1,Paul,Paul1,Paul2,Paul3}. We shall use the $\chi^{2}$
minimization technique (statistical data analysis) to get the
constraints of the parameters of MCG in RS II brane model.

\section{\bf{Observational Data Analysis Mechanism}}

From the solution (13) of MCG and defining the dimensionless
density parameters $\Omega_{m0}=\frac{\rho_{m0}}{3 H_{0}^{2}}$ and
$\Omega_{x0}=\frac{\rho_{x0}}{3 H_{0}^{2}}$ we have the expression
for Hubble parameter $H$ in terms of redshift parameter $z$ as
follows ($8\pi G=c=1$)
$$H(z)=H_{0}\left[\kappa_{4}^{2}\left\{\Omega_{x0}\left(\frac{B}{\left(1+A\right)C+B}+\frac{\left(1+A\right)C}{\left(1+A\right)C+B}\left(1+z\right)^{3\left(\alpha+1\right)\left(A+1\right)}\right)^{\frac{1}{\alpha+1}}+\Omega_{m0}\left(1+z\right)^{3(1+\omega_{m})}\right\}\right.$$
\begin{equation}
\left.\left\{1+\frac{3H_{0}^{2}}{2\lambda}\Omega_{x0}\left(\frac{B}{\left(1+A\right)C+B}+\frac{\left(1+A\right)C}{\left(1+A\right)C+B}\left(1+z\right)^{3\left(\alpha+1\right)\left(A+1\right)}\right)^{\frac{1}{\alpha+1}}+\Omega_{m0}\left(1+z\right)^{3(1+\omega_{m})}\right\}\right]^{\frac{1}{2}}
\end{equation}
From equation (14), we see that the value of $H$ depends on
$H_{0},A,B,C,\alpha,z$ so the above equation can be written as
\begin{equation}
H(z)=H_{0}E(z)
\end{equation}
where
$$E(z)=\left[\kappa_{4}^{2}\left\{\Omega_{x0}\left(\frac{B}{\left(1+A\right)C+B}+\frac{\left(1+A\right)C}{\left(1+A\right)C+B}\left(1+z\right)^{3\left(\alpha+1\right)\left(A+1\right)}\right)^{\frac{1}{\alpha+1}}+\Omega_{m0}\left(1+z\right)^{3(1+\omega_{m})}\right\}\right.$$
\begin{equation}
\left.\left\{1+\frac{3H_{0}^{2}}{2\lambda}\Omega_{x0}\left(\frac{B}{\left(1+A\right)C+B}+\frac{\left(1+A\right)C}{\left(1+A\right)C+B}\left(1+z\right)^{3\left(\alpha+1\right)\left(A+1\right)}\right)^{\frac{1}{\alpha+1}}+\Omega_{m0}\left(1+z\right)^{3(1+\omega_{m})}\right\}\right]^{\frac{1}{2}}
\end{equation}

Now $E(z)$ contains four unknown parameters $A,B,C$ and $\alpha$.
Now the relation between the two parameters will be obtained by
fixing the other two parameters and by using observational data
set. Eventually the bounds of the parameters will be obtained by
using this observational data analysis mechanism.
\[
\begin{tabular}{|c|c|c|}
\hline
  ~~~~~~$z$ ~~~~& ~~~~$H(z)$ ~~~~~& ~~~~$\sigma(z)$~~~~\\
  \hline
  0 & 73 & $\pm$ 8 \\
  0.1 & 69 & $\pm$ 12 \\
  0.17 & 83 & $\pm$ 8 \\
  0.27 & 77 & $\pm$ 14 \\
  0.4 & 95 & $\pm$ 17.4\\
  0.48& 90 & $\pm$ 60 \\
  0.88 & 97 & $\pm$ 40.4 \\
  0.9 & 117 & $\pm$ 23 \\
  1.3 & 168 & $\pm$ 17.4\\
  1.43 & 177 & $\pm$ 18.2 \\
  1.53 & 140 & $\pm$ 14\\
  1.75 & 202 & $\pm$ 40.4 \\ \hline
\end{tabular}
\]
{\bf Table 1:} The Hubble parameter $H(z)$ and the standard error
$\sigma(z)$ for different values of redshift $z$.

\subsection{Analysis with Stern ($H(z)$-$z$) Data Set}

Using observed value of Hubble parameter at different redshifts
(twelve data points) listed in observed Hubble data by
\cite{Stern} we analyze the model. The Hubble parameter $H(z)$ and
the standard error $\sigma(z)$ for different values of redshift
$z$ are given in Table 1. For this purpose we first form the
$\chi^{2}$ statistics as a sum of standard normal distribution as
follows:
\begin{equation}
{\chi}_{Stern}^{2}=\sum\frac{(H(z)-H_{obs}(z))^{2}}{\sigma^{2}(z)}
\end{equation}
where $H(z)$ and $H_{obs}(z)$ are theoretical and observational
values of Hubble parameter at different redshifts respectively and
$\sigma(z)$ is the corresponding error for the particular
observation given in table 1. Here, $H_{obs}$ is a nuisance
parameter and can be safely marginalized. We consider the present
value of Hubble parameter $H_{0}$ = 72 $\pm$ 8 Kms$^{-1}$
Mpc$^{-1}$ and a fixed prior distribution. Here we shall determine
the parameters $A,B,C$ and $\alpha$ from minimizing the above
distribution ${\chi}_{Stern}^{2}$. Fixing the two parameters
$C,\alpha$, the relation between the other parameters $A,B$ can be
determined by the observational data. The probability distribution
function in terms of the parameters $A,B,C$ and $\alpha$ can be
written as
\begin{equation}
L= \int e^{-\frac{1}{2}{\chi}_{Stern}^{2}}P(H_{0})dH_{0}
\end{equation}
where $P(H_{0})$ is the prior distribution function for $H_{0}$.
We now plot the graph for different confidence levels. In early
stage the Chaplygin Gas follow the equation of state $P=A\rho$
where $A\le 1$. So, as per our theoretical model the two
parameters should satisfy the two inequalities $A\le 1$ and $B>0$.
Now our best fit analysis with Stern observational data support
the theoretical range of the parameters. The 66\% (solid, blue),
90\% (dashed, red) and 99\% (dashed, black) contours are plotted
in figures 1, 2 and 4 for $\alpha=0.5$ and $A=1,1/3,-1/3$. The
best fit values of $B$ and $C$ are tabulated in Table 2.
\[
\begin{tabular}{|c|c|c|c|}
\hline
  ~~~~~~$A$ ~~~~~& ~~~~~~~$B$ ~~~~~~~~& ~~~$C$~~~~~&~~~~~$\chi^{2}_{min}$~~~~~~\\
  \hline
  $~~1$ & 0.5078000 & 0.114942 & 30.3789 \\
  $~~\frac{1}{3}$ & 0.0515799 & 0.764833 & 18.3400 \\
  $-\frac{1}{3}$ & 0.0226551 & 0.421103 & 8.5225 \\
   \hline
\end{tabular}
\]
{\bf Table 2:} $H(z)$-$z$ (Stern): The best fit values of $B$, $C$
and the minimum values of $\chi^{2}$ for different values of $A$.
\begin{figure}
\includegraphics[height=2in]{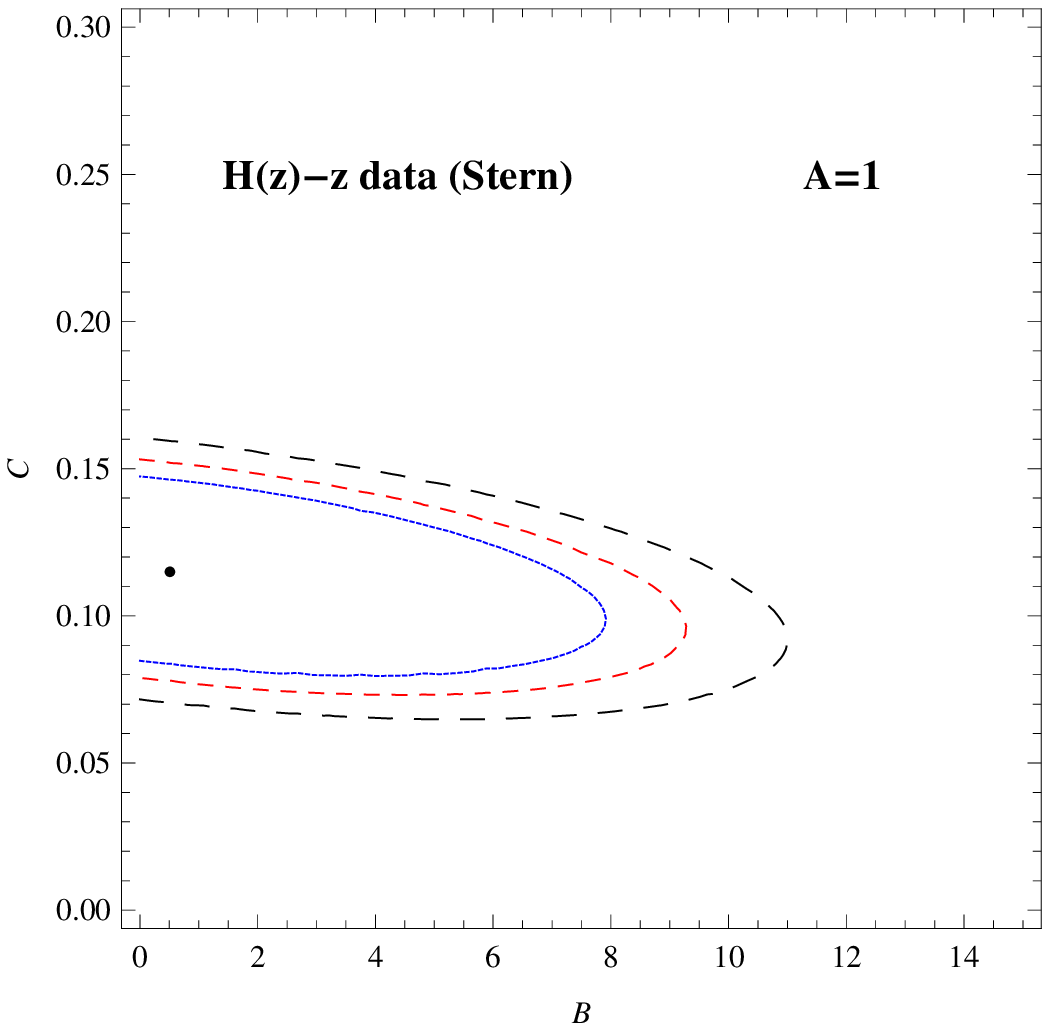}~~
\includegraphics[height=2in]{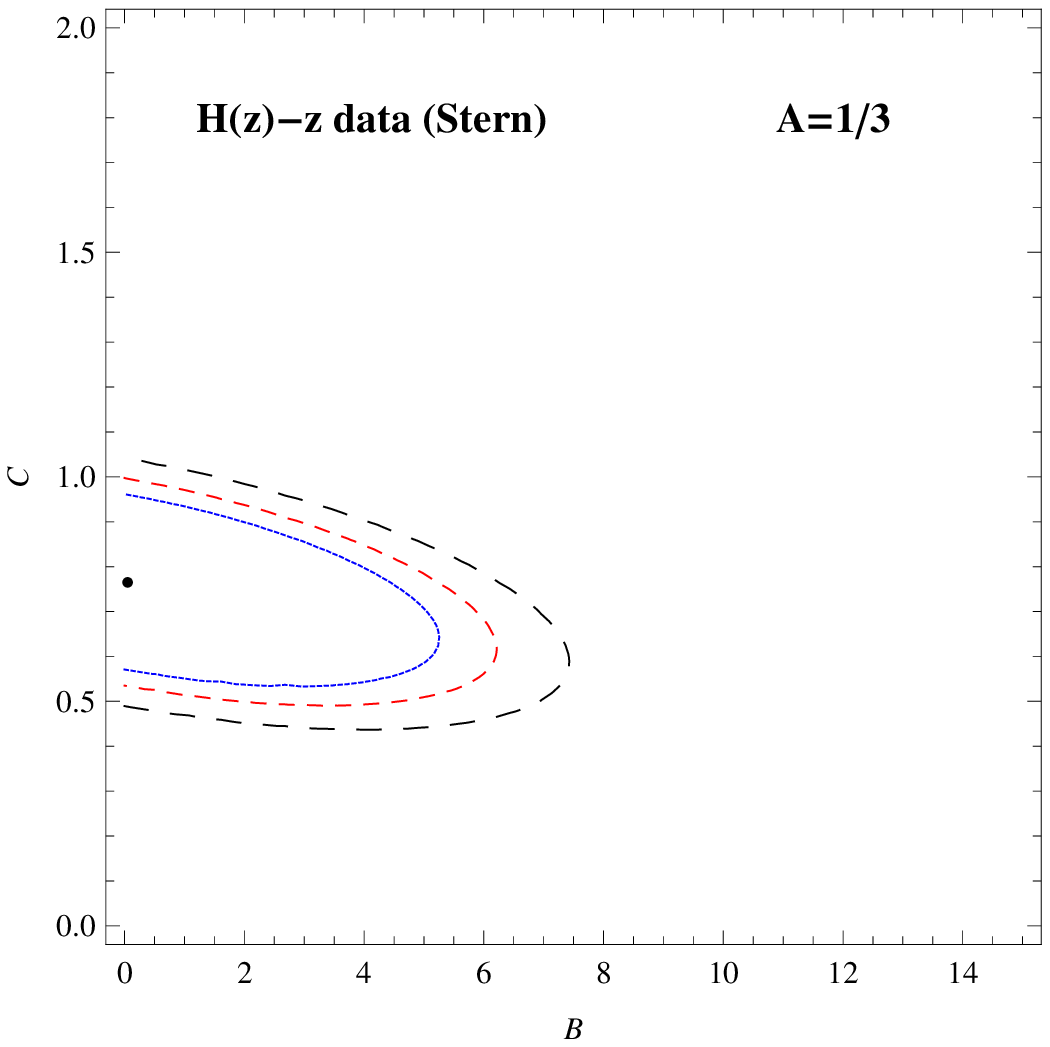}~~
\includegraphics[height=2in]{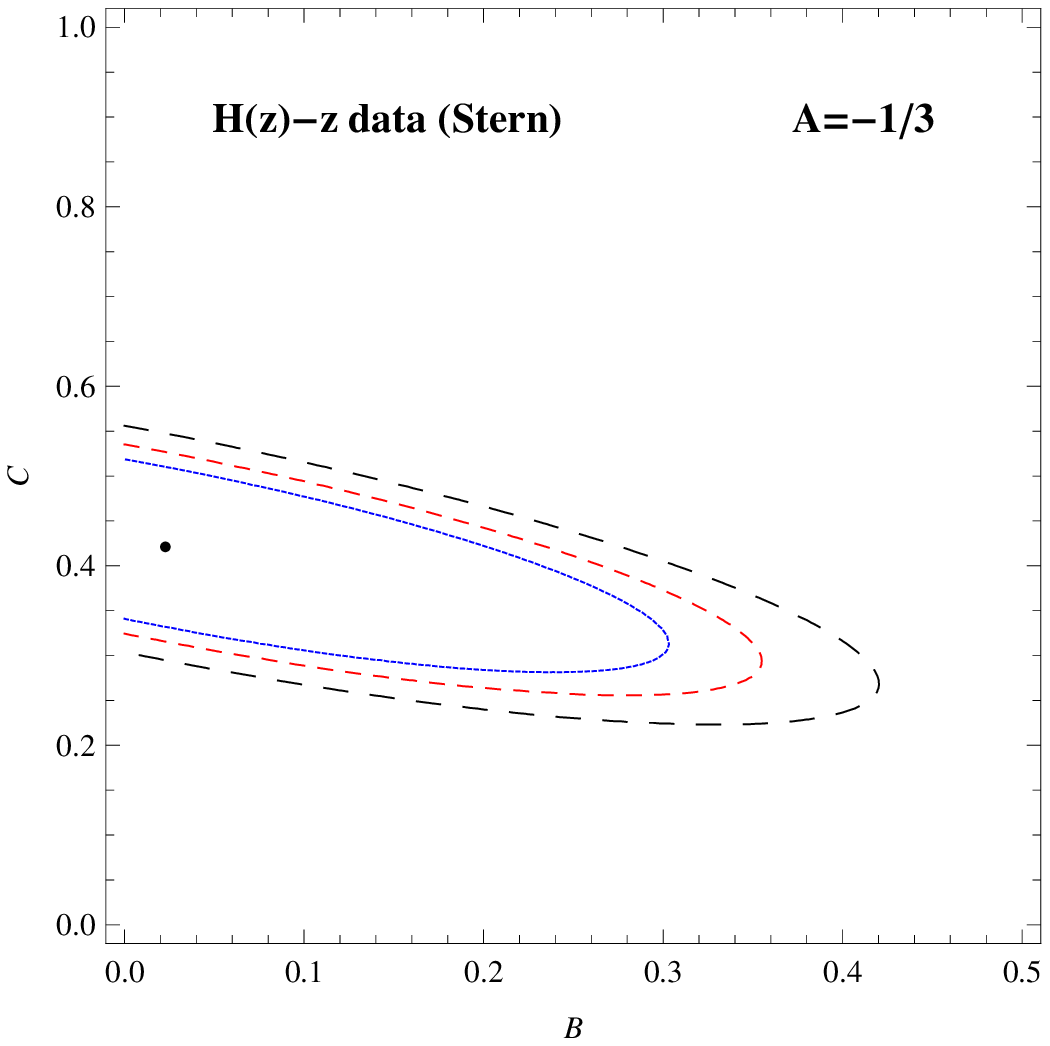}\\
\vspace{2mm}
~~~~~~~~~~~~~~~~Fig.1~~~~~~~~~~~~~~~~~~~~~~~~~~~~~~~~~~~~~Fig.2~~~~~~~~~~~~~~~~~~~~~~~~~~~~~~~~~Fig.3
\vspace{1cm}

Fig.1 shows that the variation of $C$ with $B$ for $\alpha=
0.000001$, $\Omega_{m0}=0.0013, \Omega_{x0}=0.3688$ with $A=1$ for
different confidence levels. Fig.2 shows that the variation of $C$
with $B$ for $\alpha= 0.000001$, $\Omega_{m0}=0.0012,
\Omega_{x0}=0.8035$ with $A=1/3$ for different confidence levels.
Fig.3 shows that the variation of $C$ with $B$ for $\alpha=
0.000001$, $\Omega_{m0}=0.000064, \Omega_{x0}=0.4551$ with
$A=-1/3$ for different confidence levels. The contours are plotted
for 66\% (solid, blue), 90\% (dashed, red) and 99\% (dashed,
black) confidence level in these figures for the $H(z)$-$z$
(Stern) data analysis. \vspace{1cm}
\end{figure}

\subsection{Joint Analysis with Stern $+$ BAO Data Sets}

The method of joint analysis, the Baryon Acoustic Oscillation
(BAO) peak parameter value has been proposed by \cite{Eisenstein}
and we shall use their approach. Sloan Digital Sky Survey (SDSS)
survey  is one of the first redshift survey by which the BAO
signal has been directly detected at a scale $\sim$ 100 MPc. The
said analysis is actually the combination of angular diameter
distance and Hubble parameter at that redshift. This analysis is
independent of the measurement of $H_{0}$ and not containing any
particular dark energy. Here we examine the parameters $B$ and $C$
for Chaplygin gas model from the measurements of the BAO peak for
low redshift (with range $0<z<0.35$) using standard $\chi^{2}$
analysis. The error is corresponding to the standard deviation,
where we consider Gaussian distribution. Low-redshift distance
measurements is a lightly dependent on different cosmological
parameters, the equation of state of dark energy and have the
ability to measure the Hubble constant $H_{0}$ directly. The BAO
peak parameter may be defined by
\begin{equation}
{\cal
A}=\frac{\sqrt{\Omega_{m}}}{E(z_{1})^{1/3}}\left(\frac{1}{z_{1}}~\int_{0}^{z_{1}}
\frac{dz}{E(z)}\right)^{2/3}
\end{equation}
Here $E(z)=H(z)/H_{0}$ is the normalized Hubble parameter, the
redshift $z_{1}=0.35$ is the typical redshift of the SDSS sample
and the integration term is the dimensionless comoving distance to
the to the redshift $z_{1}$ The value of the parameter ${\cal A}$
for the flat model of the universe is given by ${\cal A}=0.469\pm
0.017$ using SDSS data \cite{Eisenstein} from luminous red
galaxies survey. Now the $\chi^{2}$ function for the BAO
measurement can be written as
\begin{equation}
\chi^{2}_{BAO}=\frac{({\cal A}-0.469)^{2}}{(0.017)^{2}}
\end{equation}

Now the total joint data analysis (Stern+BAO) for the $\chi^{2}$
function may be defined by
\begin{equation}
\chi^{2}_{total}=\chi^{2}_{Stern}+\chi^{2}_{BAO}
\end{equation}

According to our analysis the joint scheme gives the best fit
values of $B$ and $C$ in Table 3. Finally we draw the contours $A$
vs $B$ for the 66\% (solid, blue), 90\% (dashed, red) and 99\%
(dashed, black) confidence limits depicted in figures $4-6$ for $\alpha=0.5$ and $A=1,1/3,-1/3$.\\

\[
\begin{tabular}{|c|c|c|c|}
\hline
  ~~~~~~$A$ ~~~~~& ~~~~~~~$B$ ~~~~~~~~& ~~~$C$~~~~~&~~~~~$\chi^{2}_{min}$~~~~~~\\
  \hline
   $~~1$ & 0.960975 & 0.0242341 & 789.179 \\
 $~~\frac{1}{3}$ & 0.676543 & 0.0553636 & 775.823  \\
 $-\frac{1}{3}$ & 0.0221079 & 0.421197 & 769.474 \\
   \hline
\end{tabular}
\]
{\bf Table 3:} $H(z)$-$z$ (Stern) + BAO : The best fit values of
$B$, $C$ and the minimum values of $\chi^{2}$ for different values
of $A$.

\begin{figure}
\includegraphics[height=2in]{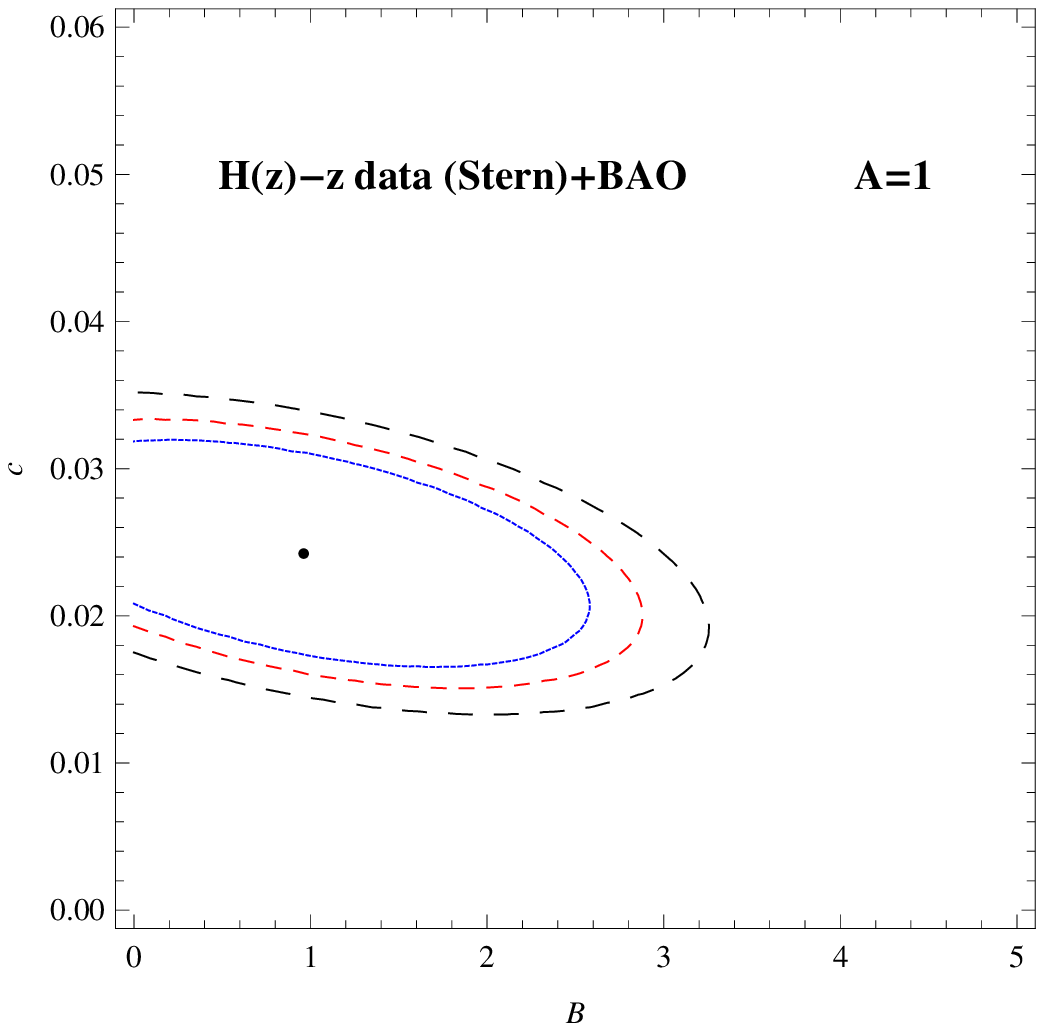}~~
\includegraphics[height=2in]{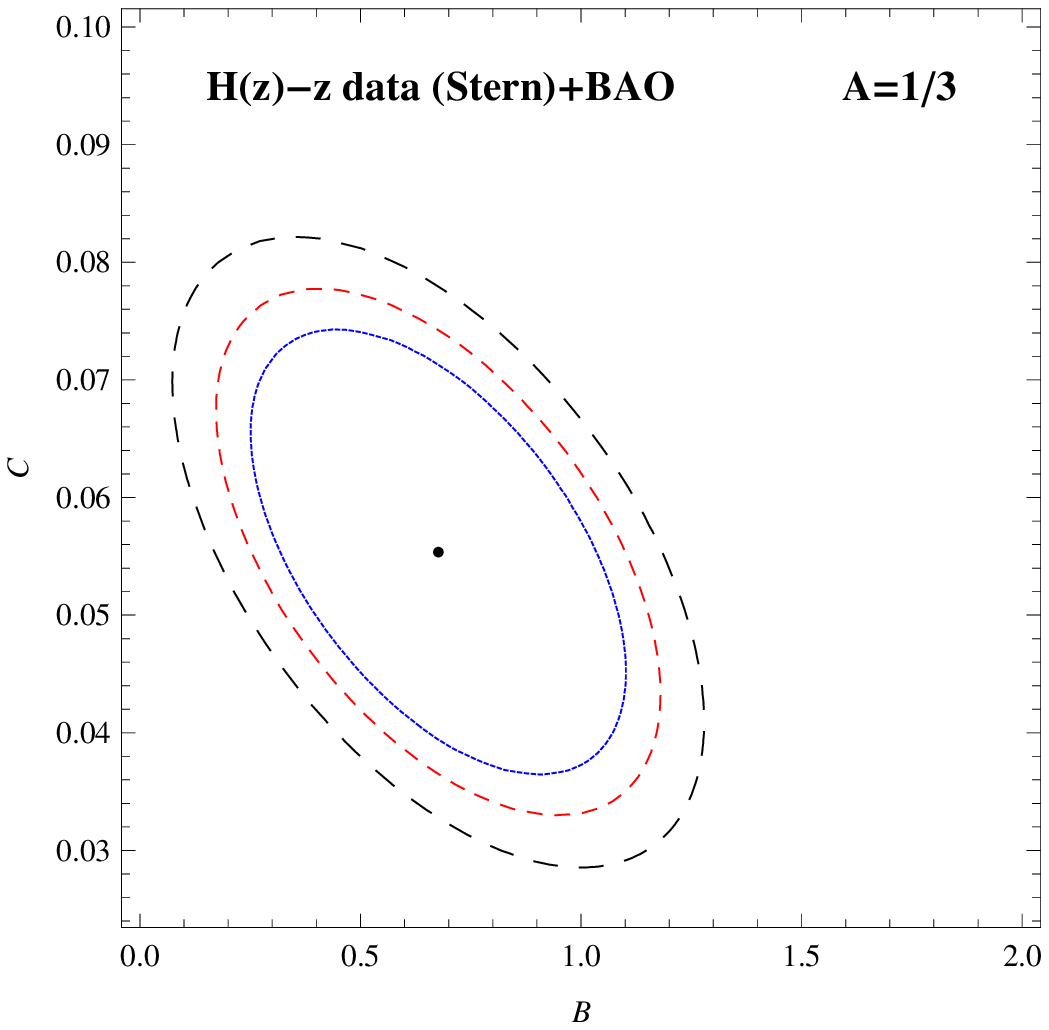}~~
\includegraphics[height=2in]{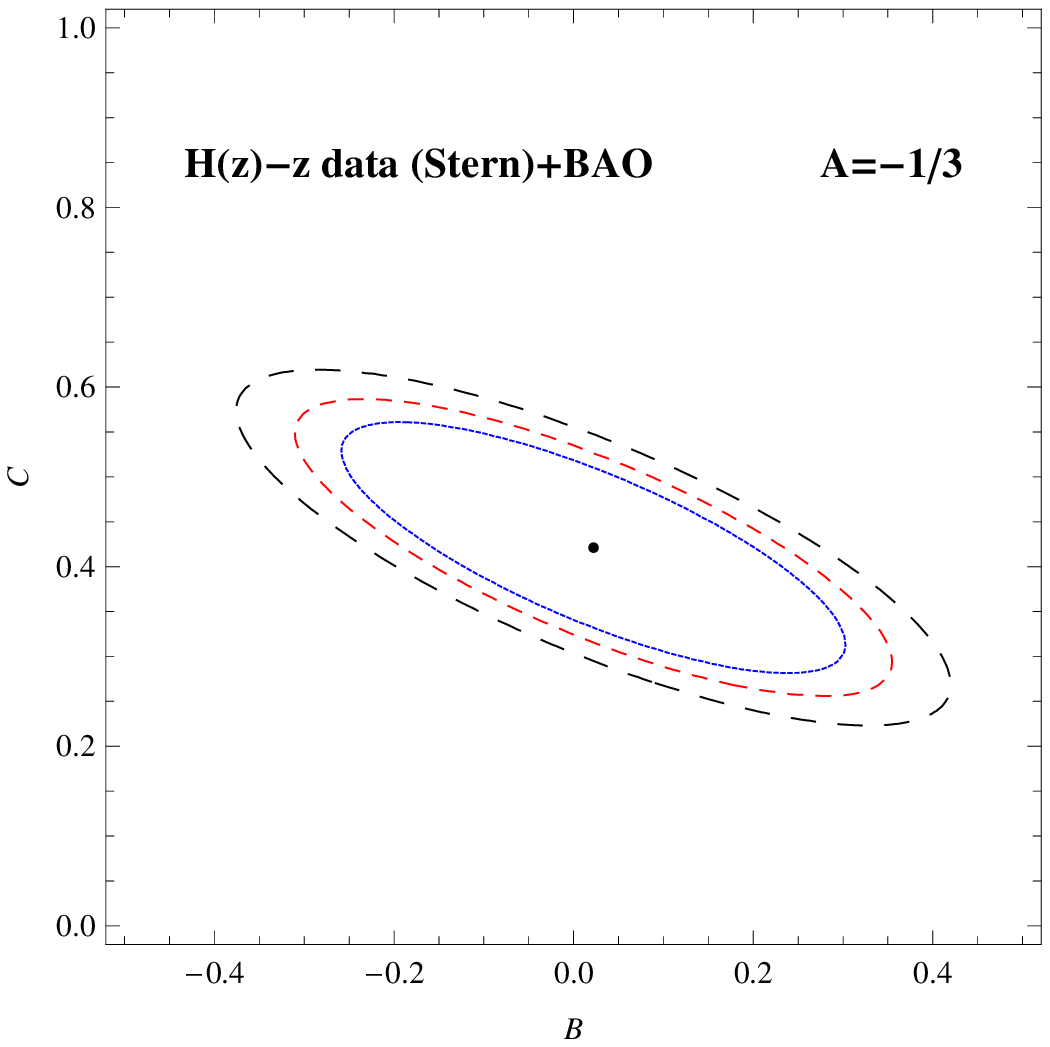}\\
\vspace{2mm}
~~~~~~~~~~~~~~~~Fig.4~~~~~~~~~~~~~~~~~~~~~~~~~~~~~~~~~~~~~Fig.5~~~~~~~~~~~~~~~~~~~~~~~~~~~~~~~~~Fig.6
\vspace{1cm}

The contours are drawn for 66\% (solid, blue), 90\% (dashed, red)
and 99\% (dashed, black) confidence levels for the $H(z)$-$z$+BAO
joint analysis. Fig.4 shows the variations of $C$ against $B$ for
$\alpha=0.000001, \Omega_{m0}=0.01, \Omega_{x0}=0.5091$ with
$A=1$. Fig.5 shows the variations of $C$ against $B$ for
$\alpha=0.000001, \Omega_{m0}=0.01, \Omega_{x0}=0.5627$ with
$A=1/3$. Fig.6 shows the variations of $C$ against $B$ for
$\alpha=0.000001, \Omega_{m0}=0.01, \Omega_{x0}=0.4544$ with
$A=-1/3$. \vspace{1cm}

\end{figure}

\subsection{Joint Analysis with Stern $+$ BAO $+$ CMB Data Sets}

One interesting geometrical probe of dark energy can be determined
by the angular scale of the first acoustic peak through angular
scale of the sound horizon at the surface of last scattering which
is encoded in the CMB power spectrum Cosmic Microwave Background
(CMB) shift parameter is defined by
\cite{Bond,Efstathiou,Nessaeris}. It is not sensitive with respect
to perturbations but are suitable to constrain model parameter.
The CMB power spectrum first peak is the shift parameter which is
given by

\begin{equation}
{\cal R}=\sqrt{\Omega_{m}} \int_{0}^{z_{2}} \frac{dz}{E(z)}
\end{equation}

\begin{figure}
\includegraphics[height=2in]{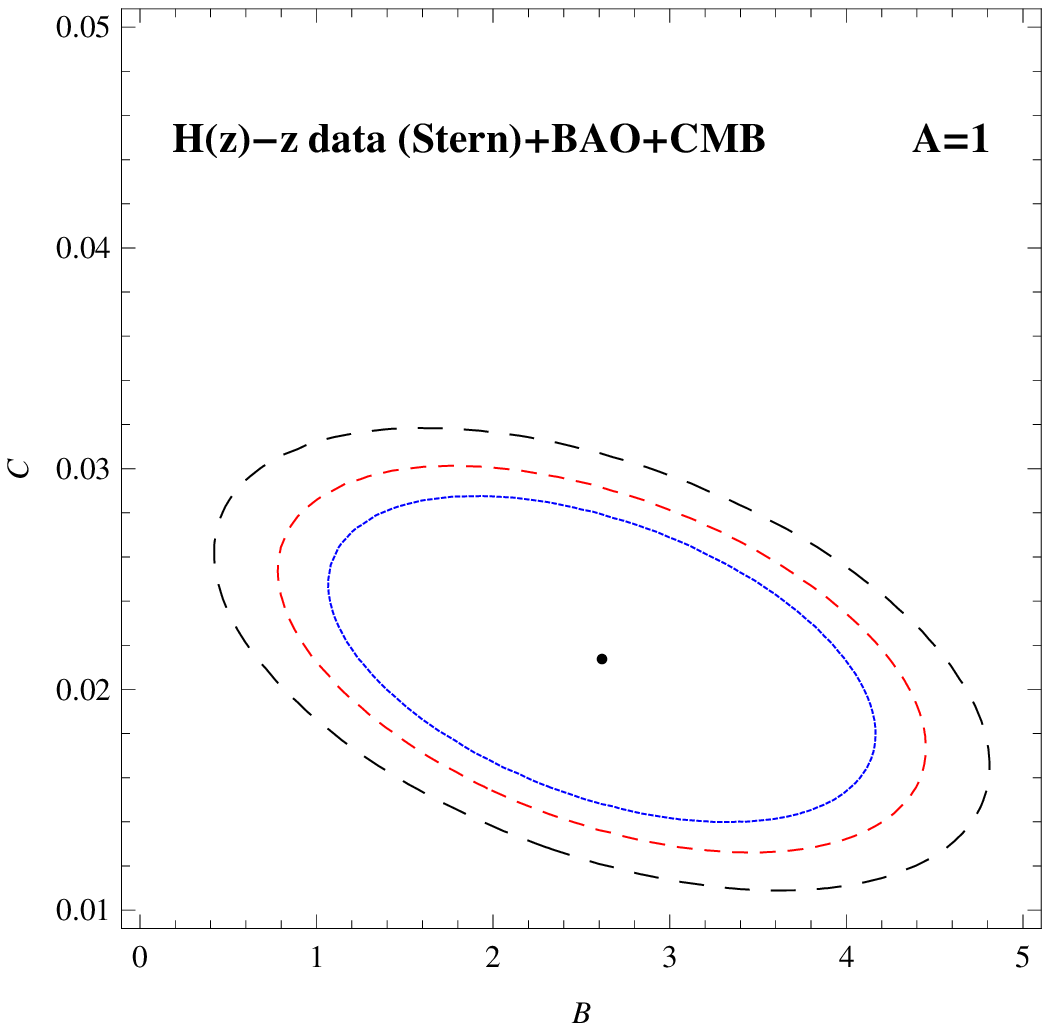}~~
\includegraphics[height=2in]{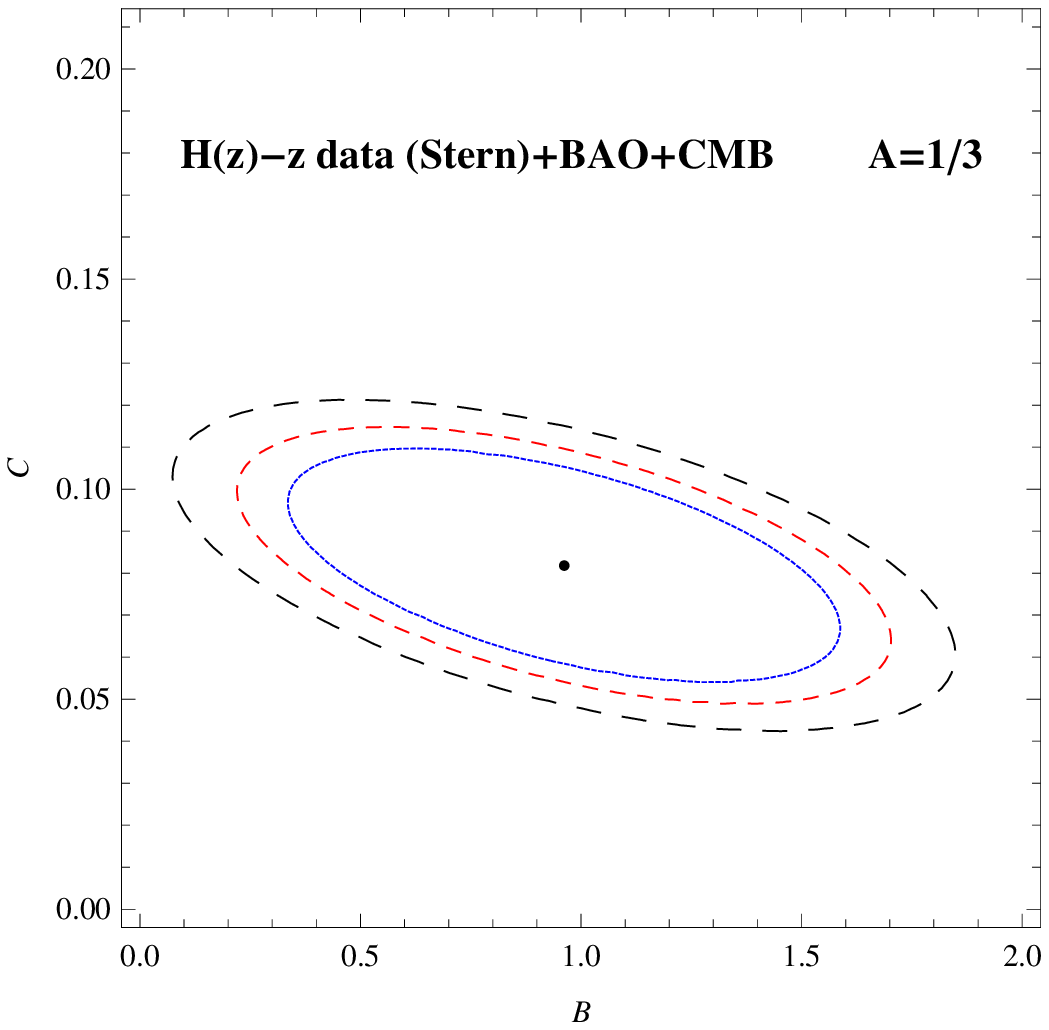}~~
\includegraphics[height=2in]{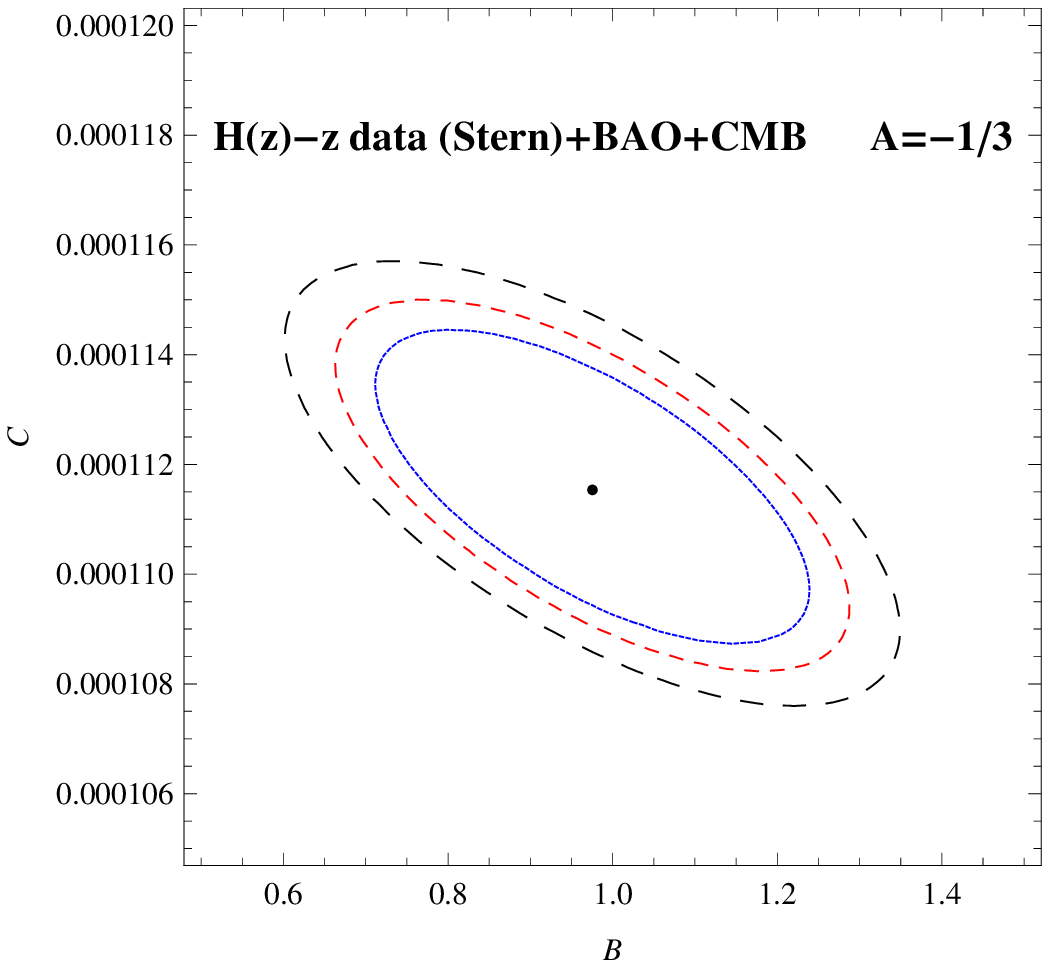}\\
\vspace{2mm}
~~~~~~~~~~~~~~~~Fig.7~~~~~~~~~~~~~~~~~~~~~~~~~~~~~~~~~~~~~Fig.8~~~~~~~~~~~~~~~~~~~~~~~~~~~~~~~~~Fig.9\\
\vspace{1cm}

The contours are drawn for 66\% (solid, blue), 90\% (dashed, red)
and 99\% (dashed, black) confidence levels for the
$H(z)$-$z$+BAO+CMB joint analysis. Fig.7 shows the variations of
$C$ against $B$ for $\alpha=0.0001, \Omega_{m0}=0.01,
\Omega_{x0}=1.3294$ with $A=1$. Fig.8 shows the variations of $C$
against $B$ for $\alpha=0.0001, \Omega_{m0}=0.01,
\Omega_{x0}=0.8033$ with $A=1/3$. Fig.9 shows the variations of
$C$ against $B$ for $\alpha=0.0001, \Omega_{m0}=0.01,
\Omega_{x0}=1.4631$ with $A=-1/3$. \vspace{1mm}

\end{figure}

where $z_{2}$ is the value of redshift at the last scattering
surface. From WMAP7 data of the work of Komatsu et al
\cite{Komatsu} the value of the parameter has obtained as ${\cal
R}=1.726\pm 0.018$ at the redshift $z=1091.3$. Now the $\chi^{2}$
function for the CMB measurement can be written as
\begin{equation}
\chi^{2}_{CMB}=\frac{({\cal R}-1.726)^{2}}{(0.018)^{2}}
\end{equation}

Now when we consider three cosmological tests together, the total
joint data analysis (Stern+BAO+CMB) for the $\chi^{2}$ function
may be defined by
\begin{equation}
\chi^{2}_{TOTAL}=\chi^{2}_{Stern}+\chi^{2}_{BAO}+\chi^{2}_{CMB}
\end{equation}
Now the best fit values of $B$ and $C$ for joint analysis of BAO
and CMB with Stern observational data support the theoretical
range of the parameters given in Table 4. The 66\% (solid, blue),
90\% (dashed, red) and 99\% (dashed, black) contours are plotted
in figures 7-9 for $\alpha=0.5$ and $A=1,1/3,-1/3$.
\[
\begin{tabular}{|c|c|c|c|}
\hline
  ~~~~~~$A$ ~~~~~& ~~~~~~~$B$ ~~~~~~~~& ~~~$C$~~~~~&~~~~~$\chi^{2}_{min}$~~~~~~\\
  \hline
  $~~1$ & 2.616200 & 0.021372500 & 9979.820 \\
  $~~\frac{1}{3}$ & 0.961993 & 0.081817900 & 9970.610 \\
  $-\frac{1}{3}$ & 0.975391 & 0.000111536 & 882.179 \\
   \hline
\end{tabular}
\]
{\bf Table 4:} $H(z)$-$z$ (Stern) + BAO + CMB : The best fit
values of $B$, $C$ and the minimum values of $\chi^{2}$ for
different values of $A$.

\begin{figure}
\vspace{2mm}
\includegraphics[scale=0.75]{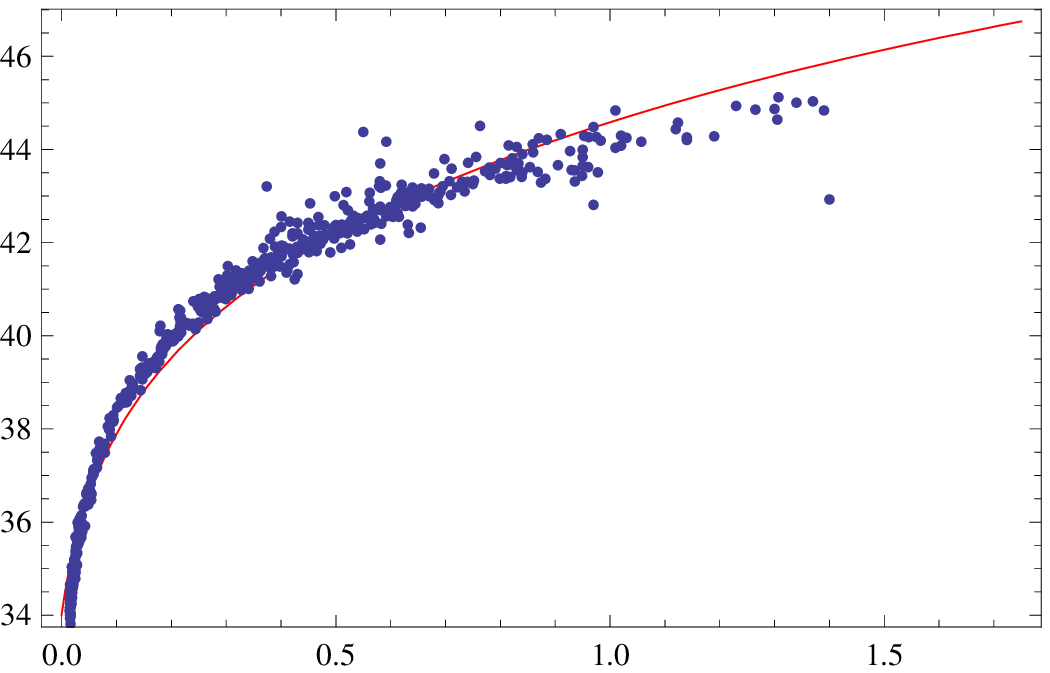}~~\\
\vspace{2mm} ~~~~~~~~~~Fig.10~~~~~~~~ \vspace{4mm}

In fig.10, $u(z)$ vs $z$ is plotted for our model (solid line) and
the Union2 sample (dotted points). \vspace{6cm}

\end{figure}

\subsection{Redshift-Magnitude Observations from Supernovae Type Ia}

The Supernova Type Ia experiments provided the main evidence for
the existence of dark energy. Since 1995, two teams of High-$z$
Supernova Search and the Supernova Cosmology Project have
discovered several type Ia supernovas at the high redshifts
\cite{Perlmutter,Perlmutter1,Riess,Riess1}. The observations
directly measure the distance modulus of a Supernovae and its
redshift $z$ \cite{Riess2,Kowalaski}. Now, take recent
observational data, including SNe Ia which consists of 557 data
points and belongs to the Union2 sample \cite{Amanullah}.

From the observations, the luminosity distance $d_{L}(z)$
determines the dark energy density and is defined by
\begin{equation}
d_{L}(z)=(1+z)H_{0}\int_{0}^{z}\frac{dz'}{H(z')}
\end{equation}
and the distance modulus (distance between absolute and apparent
luminosity of a distance object) for Supernovas is given by
\begin{equation}
\mu(z)=5\log_{10} \left[\frac{d_{L}(z)/H_{0}}{1~MPc}\right]+25
\end{equation}

The best fit of distance modulus as a function $\mu(z)$ of
redshift $z$ for our theoretical model and the Supernova Type Ia
Union2 sample are drawn in figure 10 for our best fit values of
$\alpha$, $A$, $B$ and $C$. From the curves, we see that the
theoretical MCG model in LQC is in agreement with the union2
sample data.

\section{Discussions}

In this work, we have considered the FRW universe in RS II
braneworld model filled with a combination of dark matter and dark
energy in the form of modified Chaplygin gas (MCG). MCG is one of
the candidate of unified dark matter-dark energy model. We present
the Hubble parameter in terms of the observable parameters
$\Omega_{m0}$, $\Omega_{x0}$ and $H_{0}$ with the redshift $z$ and
the other parameters like $A$, $B$, $C$ and $\alpha$. We have
chosen the observed values of $\kappa_{4}=0.1$, $\omega=-1.3$ and
$H_{0}$ = 72 Kms$^{-1}$ Mpc$^{-1}$. From Stern data set (12
points), we have obtained the bounds of the arbitrary parameters
by minimizing the $\chi^{2}$ test. Next due to joint analysis of
BAO and CMB observations, we have also obtained the best fit
values and the bounds of the parameters ($B,C$). We have plotted
the statistical confidence contour of ($B,C$) for different
confidence levels i.e., 66\%(dotted, blue), 90\%(dashed, red) and
99\%(dashed, black) confidence levels by fixing observable
parameters $\Omega_{m0}$, $\Omega_{x0}$ and $H_{0}$ and some other
parameters $A$ and $\alpha$ for Stern, Stern+BAO and Stern+BAO+CMB
data analysis.

From the Stern data,the best-fit values and bounds of the
parameters ($B,C$) are obtained for $A(=1,1/3,-1/3)$, are shown in
Table 2 and the figures 1-3 shows statistical confidence contour
for 66\%, 90\% and 99\% confidence levels. Next due to joint
analysis with Stern + BAO data, we have also obtained the best-fit
values and bounds of the parameters ($B,C$) for $A(=1,1/3,-1/3)$
and are shown in Table 3 and in figures 4-6 we have plotted the
statistical confidence contour for 66\%, 90\% and 99\% confidence
levels. After that, due to joint analysis with Stern+BAO+CMB data,
the best-fit values and bounds of the parameters ($B,C$) are found
for $A(=1,1/3,-1/3)$, are shown in Table 4 and the figures 7-9
shows statistical confidence contour for 66\%, 90\% and 99\%
confidence levels. For each case, we compare the model parameters
through the values of the parameters and by the statistical
contours. From this comparative study, one can understand the
convergence of theoretical values of the parameters to the values
of the parameters obtained from the observational data set and how
it changes for different parametric values.

Finally the distance modulus $\mu(z)$ against redshift $z$ has
been drawn in figure 10 for our theoretical model of the MCG in RS
II brane for the best fit values of the parameters and the
observed SNe Ia Union2 data sample. So the observational data sets
are perfectly consistent with our predicted theoretical MCG model
in RS II brane.

The observational study discover the constraint of allowed
composition of matter-energy by constraining the range of the
values of the parameters for a physically viable MCG in RS II
brane model. We have also verified that when $\lambda$ is large,
the best fit values of the parameters and other results of RS II
brane model in MCG coincide with the results in Einstein's gravity
\cite{Paul}. When $\lambda$ is small, the best fit values of the
parameters and the bounds of parameters spaces in different
confidence levels in RS II brane distinguished from Einstein's
gravity for MCG dark energy model. In summary, the conclusion of
this discussion suggests that even though the quantum aspect of
gravity have small effect on the observational constraint, but the
cosmological observation can put upper bounds on the magnitude of
the correction coming from quantum gravity that may be closer to
the theoretical expectation than what one would expect.\\

\section*{Acknowledgments}

The authors are thankful to IUCAA, Pune, India for warm
hospitality where a part of the work was carried out.\\


\begin{thebibliography}{99}
\bibitem[\protect\citeauthoryear{Perlmutter et al.}{1998}]{Perlmutter} Perlmutter, S. J. et al, 1998, Nature 391, 51.
\bibitem[\protect\citeauthoryear{Perlmutter et al.}{1999}]{Perlmutter1} Perlmutter, S. J. et al, 1999, Astrophys. J. 517,
565.
\bibitem[\protect\citeauthoryear{Riess et al.}{1998}]{Riess} Riess, A. G. et al., 1998, Astron. J. 116, 1009.
\bibitem[\protect\citeauthoryear{Riess et al.}{2004}]{Riess1}  Riess, A. G. et al., 2004, Astrophys. J. 607, 665.
\bibitem[\protect\citeauthoryear{Bachall et al.}{1999}]{Bachall} Bachall, N. A. et al, 1999, Science 284, 1481.
\bibitem[\protect\citeauthoryear{Tedmark et al.}{2004}]{Tedmark} Tedmark, M. et al, 2004, Phys. Rev. D 69, 103501.
\bibitem[\protect\citeauthoryear{Miller et al.}{1999}]{Miller} Miller, D. et al, 1999, Astrophys. J. 524, L1.
\bibitem[\protect\citeauthoryear{Bennet et al.}{2000}]{Bennet} Bennet, C. et al, 2000, Phys. Rev. Lett. 85, 2236.
\bibitem[\protect\citeauthoryear{Briddle et al.}{2003}]{Briddle} Briddle, S. et al, 2003, Science 299, 1532.
\bibitem[\protect\citeauthoryear{Spergel et al.}{2003}]{Spergel} Spergel, D. N. et al, 2003, Astrophys. J. Suppl. 148, 175.
\bibitem[\protect \citeauthoryear{Asthekar et al}{2011}]{Asthekar}Ashtekar, A., et al., 2011, Class. Quant. Grav., 28, 213001.
\bibitem[\protect \citeauthoryear{Cognola et al}{2009}]{Cognola}Cognola, G., et al., \textit{Phys. Rev. D}, \textbf{79}:044001(2009).
\bibitem[\protect \citeauthoryear{Chakraborty et al}{2010}]{Chakraborty2010}Chakraborty, S. and Debnath, U., \textit{Int. J. Theor. Phys.}, \textbf{24}:25(2010).
\bibitem[\protect \citeauthoryear{Ranjit et al}{2012}]{Ranjit}Ranjit, C., et al.,\textit{Int. J. Theor. Phys.}, \textbf{51}:2180(2012).
\bibitem[\protect \citeauthoryear{Brans et al}{1961}]{Brans}Brans, C. and Dicke, R.H., \textit{Phys. Rev.}, \textbf{124}:925(1961).
\bibitem[\protect \citeauthoryear{Gergely et al}{2002}]{Gergely}Gergely, L. A. and Maartens, R., \textit{Class. Quant. Grav.}, \textbf{19}:213(2002)
\bibitem[\protect\citeauthoryear{Padmanabhan}{2003}]{Paddy} Padmanabhan, T., 2003, Phys. Rept. 380, 235.
\bibitem[\protect\citeauthoryear{Sahni et al.}{2000}]{Sahni} Sahni, V. and Starobinsky, A. A., 2000, Int. J. Mod. Phys. D 9, 373.
\bibitem[\protect\citeauthoryear{Peebles et al.}{1988}]{Peebles} Peebles, P. J. E. and Ratra, B., 1988, Astrophys. J. Lett., 325,
L17.
\bibitem[\protect\citeauthoryear{Choudhury et al.}{2007}]{Paddy1} Choudhury, T. R. and Padmanabhan, T., 2007, Astron.
Astrophys. 429, 807.
\bibitem[\protect\citeauthoryear{Padmanabhan et al.}{2003}]{Paddy2} Padmanabhan, T. and Choudhury, T. R., 2003, MNRAS
344, 823.
\bibitem[\protect\citeauthoryear{Tonry et al.}{2003}]{Tonry} Tonry, J. L. et al., 2003, ApJ, 594, 1.
\bibitem[\protect\citeauthoryear{Barris et al.}{2004}]{Barris} Barris, B. J. et al., 2004, ApJ, 602, 571.
\bibitem[\protect\citeauthoryear{Amanullah et al.}{2010}]{Amanullah} Amanullah, R. et al., 2010, Astrophys. J. 716,
712.
\bibitem[\protect\citeauthoryear{Kamenshchik et al.}{2001}]{Kamenshchik} Kamenshchik, A. et al., 2001, Phys. Lett. B 511, 265
(2001).
\bibitem[\protect\citeauthoryear{Gorini et al.}{2003}]{Gorini} Gorini, V.,  Kamenshchik, A. and Moschella, U., 2003, Phys. Rev. D 67, 063509.
\bibitem[\protect\citeauthoryear{Debnath et al.}{2004}]{Debnath} Debnath, U., Banerjee, A. and Chakraborty, S., 2004, Class. Quantum Grav. 21, 5609.
\bibitem[\protect\citeauthoryear{Lu et al.}{2008}]{Lu} Lu, J. et al, 2008, Phys. Lett. B 662, 87.
\bibitem[\protect\citeauthoryear{Jun et al.}{2005}]{Jun} Dao-Jun, L. and Xin-Zhou, L., 2005, Chin. Phys. Lett., 22, 1600.
\bibitem[\protect\citeauthoryear{Rubakov et al.}{2001}]{Rubakov} Rubakov, V. A., 2001, Phys. Usp., 44, 871.
\bibitem[\protect\citeauthoryear{Maartens et al.}{2004}]{Maartens1} Maartens, R., 2004, Living Rev. Relativity, 7, 7.
\bibitem[\protect\citeauthoryear{Brax et al.}{2004}]{Brax} Brax, P. et. al., Rep. Prog.Phys., 67,
2183.
\bibitem[\protect\citeauthoryear{Randall1 et al.}{1999}]{Randall1} Randall, L., Sundrum, R.,1999, Phys. Rev. Lett., 83,
3770.
\bibitem[\protect\citeauthoryear{Randall2 et al.}{1999}]{Randall2} Randall, L., Sundrum, R., 1999, Phys. Rev. Lett., 83,
4690.
\bibitem[\protect\citeauthoryear{Ranjit et al.}{2013}]{Ranjit5}Ranjit, C., et al, 2013, Astrophys.Space Sci., DOI.:10.1007/s10509-013-1441-2.
\bibitem[\protect\citeauthoryear{Chakraborty et al.}{2012}]{Ranjit8}Chakraborty et al., 2012, Eur.Phys.J. C 72, 2101.
\bibitem[\protect\citeauthoryear{Shiromizu et al.}{2000}]{Shiromizu} Shiromizu, T., Maeda, K. and Sasaki, M., 2000, Phys. Rev.
D, 62, 024012.
\bibitem[\protect\citeauthoryear{Maeda et al.}{2000}]{Maeda} Maeda, K. and Wands, D., 2000, Phys. Rev. D, 62, 124009.
\bibitem[\protect\citeauthoryear{Sasaki et al.}{2000}]{Sasaki} Sasaki, M., Shiromizu, T. and Maeda, K., 2000, Phys. Rev.
D, 62, 024008.
\bibitem[\protect\citeauthoryear{Maartens}{2000}]{Maartens2} Maartens, R., {\it Phys. Rev. D} {\bf 62} 084023 (2000).
\bibitem[\protect\citeauthoryear{Wu et al.}{2008}]{Wu} Wu, P. and Zhang, S. N., 2008, JCAP 06, 007.
\bibitem[\protect\citeauthoryear{Chen et al.}{2008}]{Chen} Chen, S., Wang, B. and Jing, J., 2008, Phys. Rev. D 78, 123503.
\bibitem[\protect\citeauthoryear{Jamil et al.}{2011}]{jamil} Jamil, M. and Debnath, U., 2011, Astrophys Space Sci. 333, 3.
\bibitem[\protect\citeauthoryear{Fu et al.}{2008}]{Fu} Fu, X., Yu, H. and Wu, P., 2008, Phys. Rev. D 78, 063001.
\bibitem[\protect\citeauthoryear{Wu et al.}{2007}]{Wu1} Wu, P. and Yu, H., 2007, Phys. Lett. B 644, 16.
\bibitem[\protect\citeauthoryear{Thakur et al.}{2009}]{Paul} Thakur, P., Ghose, S. and Paul, B. C., 2009, Mon. Not. R. Astron. Soc. 397, 1935.
\bibitem[\protect\citeauthoryear{Paul et al.}{2011}]{Paul2} Paul, B. C., Ghose, S. and Thakur, P., arXiv:1101.1360v1 [astro-ph.CO].
\bibitem[\protect\citeauthoryear{Paul et al.}{2010}]{Paul1} Paul, B. C., Thakur, P. and Ghose, S., arXiv:1004.4256v1
[astro-ph.CO].
\bibitem[\protect\citeauthoryear{Ghose et al.}{2011}]{Paul3} Ghose, S., Thakur, P. and Paul, B. C., arXiv:1105.3303v1
[astro-ph.CO].
\bibitem[\protect\citeauthoryear{Stern et al.}{2010}]{Stern} Stern, D. et al, 2010, JCAP 1002, 008.
\bibitem[\protect\citeauthoryear{Eisenstein et al.}{2005}]{Eisenstein} Eisenstein, D. J. et al, 2005, Astrophys. J. 633, 560.
\bibitem[\protect\citeauthoryear{Bond et al.}{1997}]{Bond} Bond, J. R. et al, 1997, Mon. Not. Roy. Astron. Soc. 291, L33.
\bibitem[\protect\citeauthoryear{Efstathiou et al.}{1999}]{Efstathiou} Efstathiou, G. and Bond, J. R., 1999,
 Mon. Not. R. Astro. Soc. 304, 75.
\bibitem[\protect\citeauthoryear{Nessaeris et al.}{2007}]{Nessaeris} Nessaeris, S. and Perivolaropoulos, L., 2007, JCAP 0701, 018.
\bibitem[\protect\citeauthoryear{Komatsu et al.}{2011}]{Komatsu} Komatsu, E. et al, 2011, Astrophys. J. Suppl. 192, 18.
\bibitem[\protect\citeauthoryear{Riess et al.}{2007}]{Riess2}  Riess, A. G. et al., 2007, Astrophys. J. 659, 98.
\bibitem[\protect\citeauthoryear{Kowalaski et al.}{2008}]{Kowalaski} Kowalaski et al, 2008, Astrophys. J. 686, 749.
\end{thebibliography}
\end{document}